\documentclass{article}
\usepackage[utf8]{inputenc}
\usepackage{authblk}
\usepackage{setspace}
\usepackage[margin=1.25in]{geometry}
\usepackage{graphicx}
\graphicspath{ {./Figures/} }
\usepackage{subcaption}
\usepackage{amsmath}

\usepackage[style=ieee, 
citestyle=numeric-comp,sorting=none]{biblatex}
\title{Factor 30 pulse compression by hybrid multi-pass multi-plate spectral broadening}

\author[1*]{Marcus Seidel}
\author[1,2,3]{Prannay Balla}
\author[1]{Chen Li}
\author[4]{Gunnar Arisholm}
\author[1]{Lutz Winkelmann}
\author[1]{Ingmar Hartl}
\author[1,2,3]{Christoph M. Heyl}

\affil[1]{Deutsches Elektronen-Synchrotron DESY, Notkestrasse 85, 22607 Hamburg, Germany.}
\affil[2]{Helmholtz-Institute Jena, Fr\"{o}belstieg 3,07743 Jena, Germany.}
\affil[3]{GSI Helmholtzzentrum f\"{u}r Schwerionenforschnung GmbH, Planckstrasse 1, 64291 Darmstadt, Germany.}
\affil[4]{FFI (Norwegian Defence Research Establishment), P. O. Box 25, NO-2027 Kjeller, Norway.}
\affil[*]{Corresponding author. Email: marcus.seidel@desy.de}

\date{\today}

\begin{document}

\maketitle

\begin{abstract}
As Ultrafast laser technology advances towards ever higher peak and average powers, generating sub-50 fs pulses from laser architectures that exhibit best power-scaling capabilities remains a major challenge. Here, we present a very compact and highly robust method to compress 1.24\,ps pulses to 39\,fs by means of only a single spectral broadening stage which neither requires vacuum parts nor custom-made optics. Our approach is based on the hybridization of the multi-plate continuum and the multi-pass cell spectral broadening techniques. Their combination leads to significantly higher spectral broadening factors in bulk material than what has been reported from either method alone. Moreover, our approach efficiently suppresses adverse features of single-pass bulk spectral broadening. We use a burst mode Yb:YAG laser emitting pulses with 80\,MW peak power that are enhanced to more than 1\,GW after post-compression. With only 0.19\,\% rms pulse-to-pulse energy fluctuations, the technique exhibits excellent stability. Furthermore, we have measured state-of-the-art spectral-spatial homogeneity and good beam quality of M$^2 = 1.2$ up to a spectral broadening factor of 30. Due to the method's simplicity, compactness and scalability, it is highly attractive for turning a high-power picosecond laser into an ultrafast light source that generates pulses of only a few tens of femtoseconds duration. 
\end{abstract}

\section{Introduction}
\begin{figure}[b!]
	\centering
	\includegraphics[width=.7\linewidth]{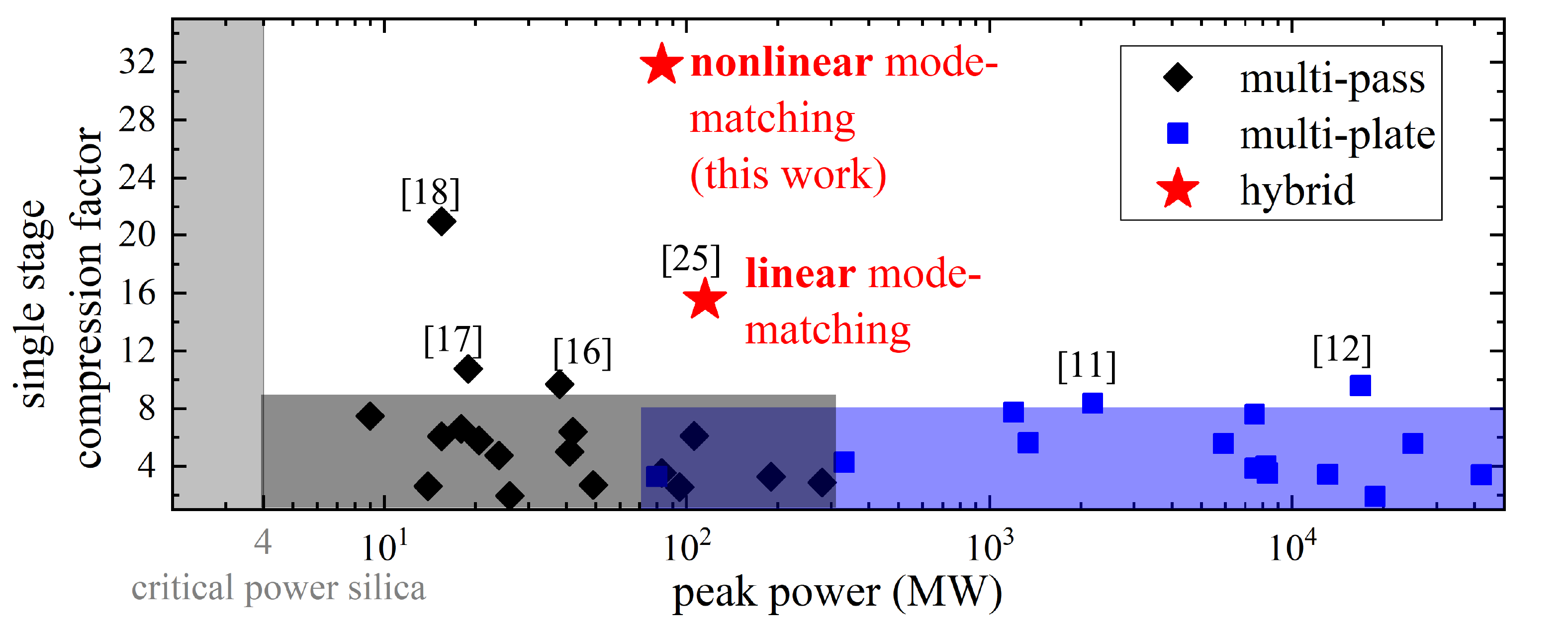}
	\caption{Overview of single stage compression factors achieved by bulk-based MPCs and the multi-plate continuum technique. The light gray and blue shaded areas enclose the typical operation regimes of the respective compression techniques. References to the experiments with the highest reported single-stage compression factors are provided.}
	\label{fig:Fig1_Overview}
\end{figure}
Femtosecond light pulses are used for a variety of applications including time-resolved fundamental science, material processing and medical applications \cite{mourou_nobel_2019}. Present ultrafast laser developments reach for increasing peak- and average powers \cite{fattahi_third-generation_2014} in order to advance applications in these fields, for instance by laser-particle acceleration \cite{albert_2020_2021}, nonlinear attosecond science \cite{orfanos_attosecond_2019} and raw-event detection utilized in coincidence spectroscopy \cite{mikaelsson_high-repetition_2021}. Today's most common high-power ultrafast light sources rely on active Yb-ions. A major downside of these lasers is the inability to directly deliver sub-100\,fs pulses. Therefore, a multitude of applications requiring shorter pulses cannot readily be addressed. To overcome this limitation, spectral broadening and subsequent pulse compression have become a thriving research topic \cite{nagy_high-energy_2021}. Different spectral broadening methods methods are commonly employed. Due to their excellent power-handling capabilities, bulk nonlinear media and multi-pass cells (MPCs) employing bulk- or gas-based nonlinear spectral broadening present a vital alternative to fibers or hollow-core capillaries. This was facilitated by the inventions of MPC spectral broadening \cite{schulte_nonlinear_2016} and multi-plate continuum generation \cite{lu_generation_2014}. In both approaches, the introduced quasi-waveguide results in much better spatial homogeneities of the broadband spectra than in bare bulk experiments which were earlier proposed as a simple method for extending optical pulse bandwidths \cite{rolland_compression_1988}. \par
Here, we present the combination of both techniques in a single, compact spectral broadening stage. We demonstrate unprecedented high pulse compression factors of more than 30 (see Fig.~\ref{fig:Fig1_Overview}), corresponding to a reduction of the FWHM pulse duration from 1240\,fs to 39\,fs. Despite strong self-phase modulation, we efficiently suppress spatial spectrum variations that result from single-pass bulk spectral broadening in the critical self-focusing regime \cite{seidel_all_2016}. For comparison, the highest pulse compression factors reported from single multi-plate continuum stages are 8.3 \cite{huang_polarization_2018} and 9.6 \cite{lu_greater_2019}. Large compression factors have been achieved with gas-filled MPCs at multi-mJ pulse energy levels \cite{kaumanns_multipass_2018,balla_postcompression_2020,kaumanns_spectral_2021}, but only a few single bulk MPC stages were reported to be able to achieve more than 10-fold compression. Those methods, however, either relied on custom-made dispersive mirrors \cite{grobmeyer_self-compression_2020} or multi-ps long input pulses being insensitive to dispersion \cite{song_generation_2021,song_generation_2021-1}. In contrast, the implementation of our method requires only off-the-shelf optics and enables strong spectral broadening by high B-integrals per pass.
Therefore, our compression scheme is a compact alternative to more complex multi-stage setups \cite{lu_greater_2019,fritsch_all-solid-state_2018,vicentini_nonlinear_2020,barbiero_broadband_2020,tsai_efficient_2019,barbiero_efficient_2021} and presents a highly attractive method to generate sub-50\,fs pulses from a high-power picosecond laser. \par

\section{Materials and Methods}
\begin{figure}[b!]
	\centering
	\includegraphics[width=.7\linewidth]{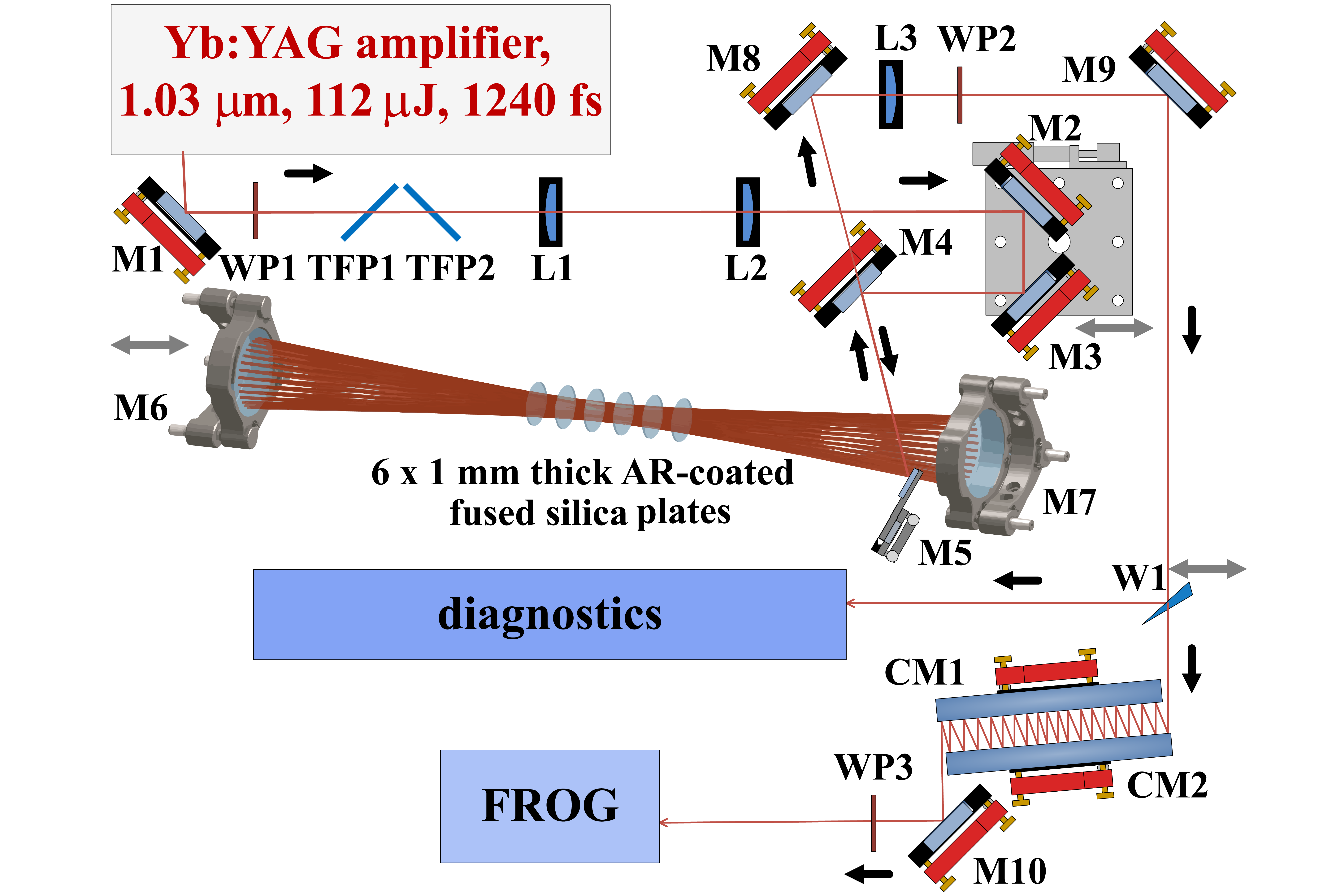}
	\caption{Spectral broadening setup. Waveplate WP1 and thin-film polarizers TFP1 and TFP2 are used for polarization cleaning of the Yb:YAG amplifier output. Mode-matching is adjusted with the lenses L1 and L2 as well as the translation stage with mirrors M2 and M3 on it. The mirror M4 is lower than the other mirrors, such that the ingoing beam hits M4 on the top and the outgoing beam is passing it. The focal length of both MPC mirrors M6 and M7 is 100\,mm. Pick-off mirror M5 has only a 4 mm vertical aperture to not clip the beam despite of the 32 roundtrips. The six 1\,mm thin Kerr media are centered in the about 385\,mm long MPC, mounted in lens tubes and spaced by 16\,mm. The lens L3 is used for collimation, the waveplate WP2 for readjusting p-polarization to match the specification of the chirped mirrors CM1 and CM2. A wedge is used for sending the beam to all diagnostics except from a FROG-device which is behind the compressor.}
	\label{fig:Fig2_Setup}
\end{figure}
\subsection{Experimental Design}
We compressed the output of a multi-stage Yb:YAG amplifier operating in 10 Hz burst mode being matched to the pulse trains of the free-electron laser FLASH at DESY \cite{li_flexible_2019,seidel_ultrafast_2021}. The system delivered pulses centered at 1030\,nm wavelength with 112\,$\mu$J energy, near Fourier-transform limited 1240\,fs pulse duration and a peak power of about 80\,MW. Frequency resolved optical gating (FROG) measurements of the pulses are provided in supplement~\ref{S:FROG}. The maximal in-burst average power with minimized pulse-to-pulse energy fluctuations was 112\,W. The corresponding 1\,MHz pulse repetition rate could be maintained over the burst duration of 800\,$\mu$s. To avoid saturation of our measurement devices, we used an acousto-optic modulator to adjust the number of pulses in the bursts and their amplitude \cite{seidel_ultrafast_2021}.\par 
Whereas a high-pressure gas-filled multi-pass cell could have been used at our peak power level \cite{lavenu_high-power_2019}, we decided to spectrally broaden the pulses in a bulk-based MPC. This obviated the need for a complex overpressure system. The setup is described in the caption of Fig.~\ref{fig:Fig2_Setup}. The employed multilyaer MPC mirrors were standard quarter-wave stacks with 200\,mm radius of curvature and 50.8\,mm diameter. All Kerr media used were stock items. In this way, the setup was kept as simple as possible. We used two focusing lenses and a translation stage for adjusting mode-matching while monitoring the beam size in the targeted focal plane inside the MPC.\par
We measured pulse durations by second-harmonic FROG with a commercial device (Mesa Photonics). In the diagnostics section of the setup, we used a compact grating spectrometer or an optical spectrum analyzer (OSA) for spectrum measurements and a Spiricon M200 M$^2$-meter for beam quality evaluation. An ANDOR Kymera 193 Czerny-Turner-type spectrograph with a 10\,$\mu$m horizontal entrance slit and a grating in the $2f$-plane with 500 lines/mm was used for measuring spectral-spatial homogeneity. We placed a FLIR grasshopper CCD camera in the $4f$ plane that recorded the dispersed spectrum along the horizontal axis and a line-out of the beam along the vertical axis. The wavelength axis was calibrated by comparison to a spectrum recorded with the compact grating spectrometer. The spectral response of the spectrograph was not determined. The beam profiles were measured with the same camera, which had 3.69$\,\mu$m $\times$ 3.69$\,\mu$m pixel size.

\subsection{Simulations}
\label{sec:meth_sim}
We support our experimental data by nonlinear beam propagation simulations. We employed the SISYFOS code \cite{arisholm_simulation_2012} which was previously used for bulk spectral broadening simulations \cite{seidel_all_2016}. The Herriott-cell shown in Fig.~\ref{fig:Fig2_Setup} was modeled under the following approximations: the beam was always on the optical axis of the system, a fundamental Gaussian beam ($M^2 = 1$) was propagated, reflectivity and dispersion of the MPC mirrors as well as transmission of the anti-reflective (AR) coatings were taken from theoretical data provided by the suppliers. We included the Kerr nonlinearity of air but we neglected the non-instantaneous Raman contributions to the nonlinearities of fused silica and air. Self-steeping was implicitly considered by solving the wave equation in the frequency domain.\par
Test simulations with both transverse dimensions and a single transverse dimension, exploiting cylindrical symmetry, led to the same outcome.  Therefore, we used the computationally less expensive reduced spatial dimensionality with 512 radial grid points. The time and frequency grid, respectively, consisted of 2048 points to properly model the narrow initial and the broad final spectrum.\par
The Kerr media's nonlinear refractive index, $n^\text{sim}_2 = 2.5\cdot10^{-16}$cm$^2$/W, used in simulations was deduced from a comparison between simulation and an MPC experiment with factor ten spectral broadening. It is in good agreement with literature values \cite{milam_review_1998}. We define the spectral broadening factor as the ratio between initial and broadened spectrum transform-limited FWHM pulse durations.

\section{Results}
\subsection{Hybrid multi-plate, multi-pass approach}
\begin{figure}[b!]
	\centering
	\includegraphics[width=.9\linewidth]{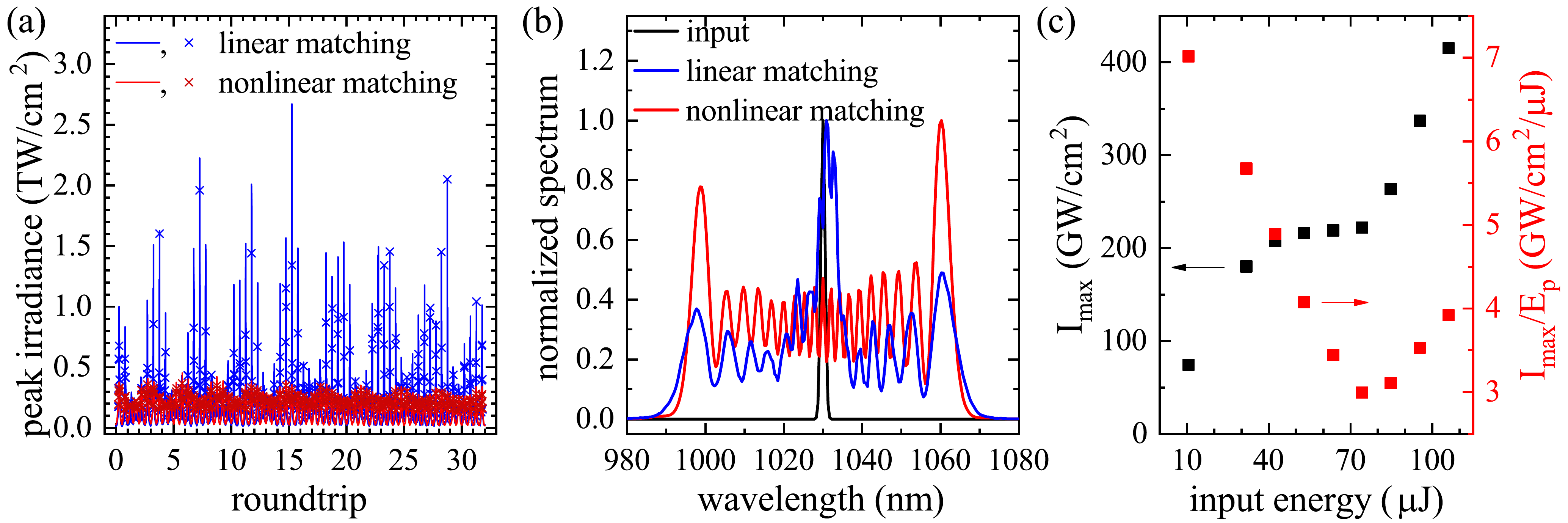}
	\caption{\textbf{(a)} Comparison of simulated peak irradiances inside the MPC shown in Fig.~\ref{fig:Fig2_Setup}(a) for 100\,$\mu$J, 1250\,fs Gaussian input pulses in case of linear (blue line) and nonlinear (red line) mode-matching. The crosses indicate the locations of the silica plates. Nonlinear mode-matching significantly reduces the maximal peak irradiance. Residual peak irradiance oscillations are caused by the imprecision of our analytical model. Best mode-matching is accomplished for 75\,$\mu$J input pulses according to (c). \textbf{(b)} Output spectra after 32 roundtrips corresponding to irradiance evolution from (a) and comparison to the input spectrum. \textbf{(c)} Simulated dependence of maximal peak irradiance in the Kerr media (I$_\text{max}$) on input pulse energy (E$_\text{p}$) for a nonlinear mode-matching point of about 75\,$\mu$J where I$_\text{max}$/E$_\text{p}$ reaches its minimum. The monotonic increase of I$_\text{max}$ obviates the need for modifying the input beam during energy ramp-up in a nonlinear mode-matching setting.}
	\label{fig:Fig3_Simulation}
\end{figure}
We conducted initial experiments with single silica plates of 3\,mm to 10\,mm thickness inside an approximately 350\,mm long MPC. The thick Kerr media exhibited two drawbacks: First, in addition to spectral broadening, we observed an accumulation of spectral power near the fundamental wavelength of 1030\,nm with increasing input pulse energy, indicating a power-dependent mode-mismatch. We even damaged the Kerr medium after extending the MPC to 380\,mm length. Second, we measured the spectral phase of pulses compressed by approximately a factor of ten and retrieved a characteristic phase kink near the fundamental wavelength which is known from bulk spectral broadening \cite{seidel_all_2016}. We attribute these observations to nonlinear beam reshaping inside the Kerr medium. This is supported by a set of free beam nonlinear propagation simulations that yield a spectral broadening factor of 1.3. The simulations are described in supplement~\ref{S:single_pass_enhancement}. We found that an intense beam propagating through a short medium undergoes significantly less deformation compared to a less intense beam propagating through a longer medium. This indicates that limiting the B-integral per pass, as proposed in the original MPC spectral broadening patent \cite{andreas_vernaleken_method_2015}, may not be necessary to suppress nonlinear space-time coupling. The B-integral per Kerr medium in multi-plate continuum generation was clearly higher than in bulk MPC experiments. Nonetheless, good spatial-spectral homogeneity was reported \cite{lu_greater_2019,cheng_supercontinuum_2016}. Consequently, we used this approach and distributed the nonlinear phase accumulation among a sequence of multiple thin plates inside the Herriott-cell.\par 
Furthermore, Kerr lensing plays a crucial role in multi-plate continuum generation because it establishes a quasi-waveguide \cite{seidel_all_2016,cheng_supercontinuum_2016}. We also included the self-focusing effect in the mode-matching calculations (see supplement~\ref{S:mode-match}) of our hybrid multi-plate, multi-pass scheme. We refer to this as nonlinear mode-matching in bulk MPCs. Herriott pattern, cell length and beam curvature at the mirrors correspond to the linear mode-matching calculations but including the Kerr effect yields smaller spot sizes at the mirrors and larger ones in the cell center (see Fig.~\ref{fig:S_mode-match}).\par
 Fig.~\ref{fig:Fig3_Simulation}(a) shows the simulated peak irradiance inside the MPC illustrated in Fig.~\ref{fig:Fig2_Setup}. The difference between linear and nonlinear mode-matching is striking. Whereas for linear mode-matching the peak irradiance in the Kerr media reaches up to 2\,TW/cm$^2$, which would cause coating damage in an experiment, the peak irradiance stays below 500\,GW/cm$^2$ for nonlinear mode-matching, preventing self-focusing induced damage. Consequently, high peak intensities in the nonlinear media can be obtained over all passes by accounting for Kerr lensing. This, in turn, enables higher spectral broadening factors from a single stage. A recent theory paper has also pointed out the importance of nonlinear mode-matching in gas-filled MPCs that operate in the sub-critical focusing regime \cite{hanna_nonlinear_2021}. The irradiance enhancement predicted for the bulk MPC studied here is, however, more than 10 times higher than in the numerical example of ref.~\cite{hanna_nonlinear_2021}.\par 
 Our simulations indicate that the shape of the nonlinearly broadened spectra indicate the quality of mode-matching (see Fig.~\ref{fig:Fig3_Simulation}(b)). With nonlinear mode-matching, the outermost spectral lobes exhibit the highest spectral power. By contrast, excessive spectral power is concentrated near the fundamental wavelength if only the linear cavity is mode-matched. This is in agreement with our experimental observations.  We note that a distinct peak around 1030\,nm is also caused by unbroadened input pulse pedestals or pre/post pulses. This peak is, however, independent of power launched into the MPC. \par

\subsection{Spectral broadening and pulse compression experiments}
 In order to verify the occurrence of self-focusing and to determine an experimental nonlinear refractive index, we conducted Z-scan-type measurements inside the MPC (see supplement~\ref{S:Z-scan}). We used the result to calculate the beam parameters for non-linear mode-matching and aligned the MPC correspondingly (see supplement~\ref{S:mode-match} for details). We did not adjust any optics while increasing the input power but expected the laser beam to be matched to the Herriott-cell at full input energy. This is possible because the maximal peak irradiances in the Kerr media monotonically increase with input pulse energy according to our simulations shown in Fig.~\ref{fig:Fig3_Simulation}(c). \par
 
\begin{figure}[t]
 	\centering
 	\includegraphics[width=.7\linewidth]{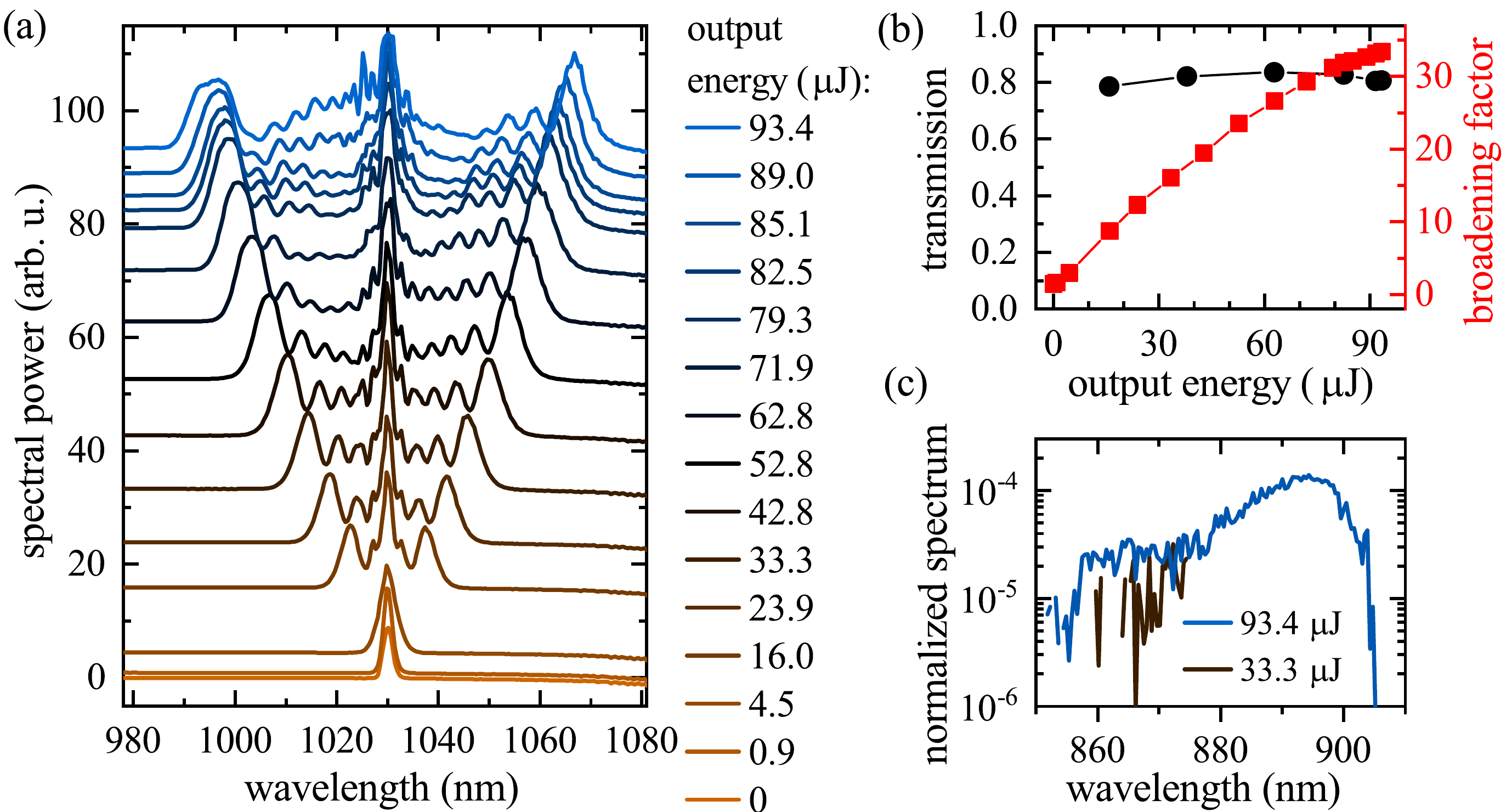}
 	\raisebox{-1mm}{\includegraphics[width=.29\linewidth]{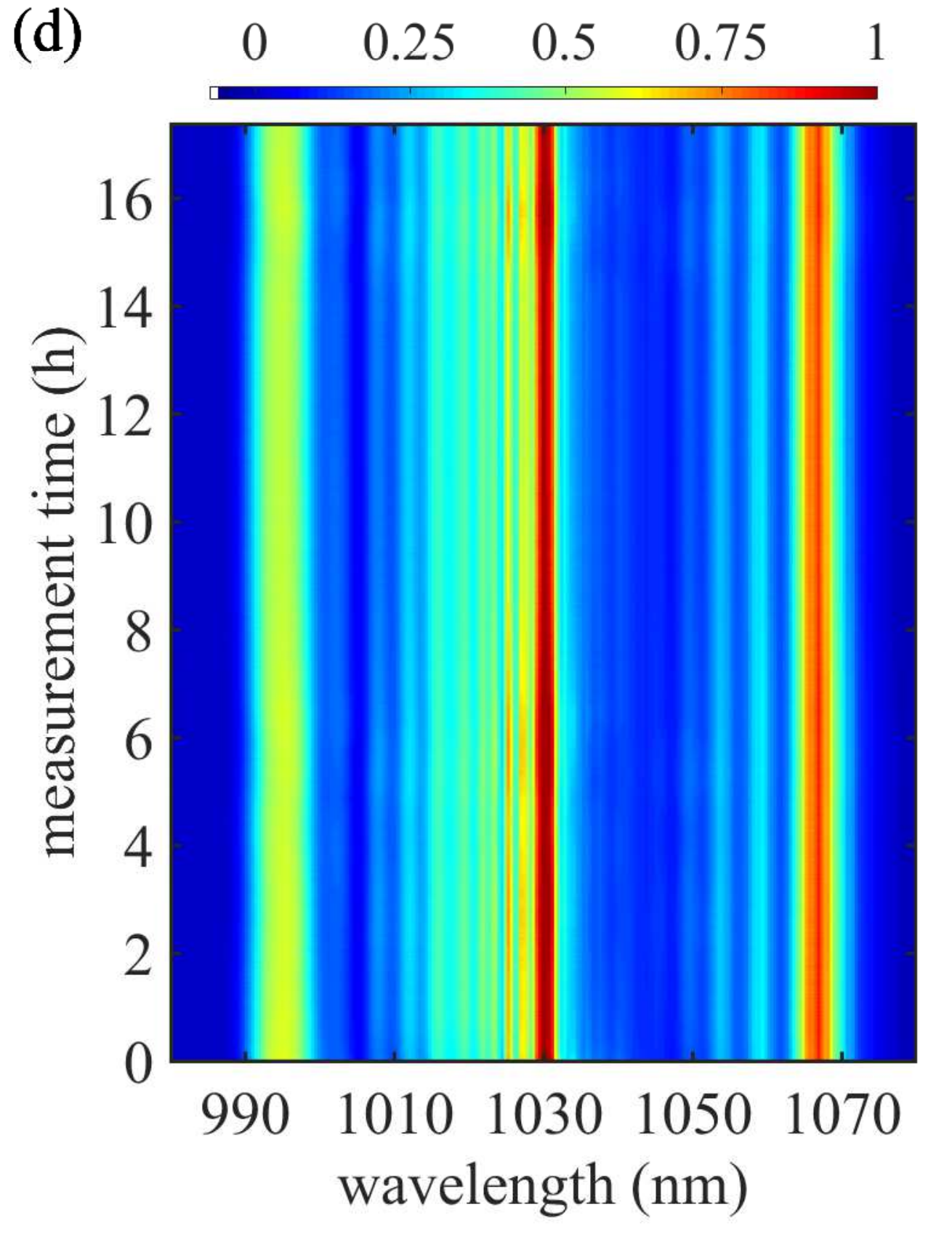}}
 	\caption{\textbf{(a)} Broadened spectra after 32 roundtrips in a 385\,mm long MPC detected with a response function corrected grating spectrometer. The spectra are offset by the output energy measured with an energy meter and corrected by the wedge W1 transmission. \textbf{(b)} Corresponding, spectral broadening factors as a function of output pulse energy. The transform-limited pulse duration at full power is more than 30 times shorter than at the input. The transmission is constantly about 80\,\%. \textbf{(c)} Spectral side-bands, becoming detectable for more than 30\,$\mu$J output energy. These are excited by four-wave mixing as qualitatively predicted by simulations. The spectra are normalized to the maximum of the respective self-phase modulated part near 1030\,nm. \textbf{(d)} Spectrum stability measured over 17 hours with one recording per second.}
 	\label{fig:Fig4_Broadening}
 \end{figure}

 Fig~\ref{fig:Fig4_Broadening}(a) shows a set of measured broadened spectra that were recorded behind the MPC while the laser pulse energy was increased. The derived dependence of the spectral broadening factor on the measured output pulse energy is plotted in Fig~\ref{fig:Fig4_Broadening}(b). The dispersion of the six Kerr plates yielded saturation of self-phase modulation at high input powers. Sub-40\,fs Fourier-transform limits were reached at about 75\,$\mu$J output pulse energy. The highest achieved spectral broadening factor was 33, corresponding to a reduction of bandwidth-limited pulse duration from 1.2\,ps to 36\,fs. We note that the Fourier tranform-limit was independent of the number of pulses per laser burst, i.e., the same spectral broadening characteristics were also obtained with 112\,W average power over the maximal 800\,$\mu$s burst duration. The measured spectra in Fig.~\ref{fig:Fig4_Broadening}(a) exhibit a central spectral peak at both low and high output energies. It was caused by a pedestal of the input pulses which we measured by FROG (see supplement~\ref{S:FROG}). Aside from that, the spectral shape agrees well with the simulated one in Fig.~\ref{fig:Fig3_Simulation}(b) up to approximately 50\,$\mu$J output energy. For higher energies, the spectral modulations flatten out at parts of the spectrum and the symmetry with respect to the fundamental wavelength decreases. We attribute this to the spurious four-wave mixing which was qualitatively predicted in simulations (see supplement~\ref{S:FWM}) and was experimentally confirmed as shown in  Fig.~\ref{fig:Fig4_Broadening}(c). The generation of spectral side-bands was previously observed in gas-filled MPC experiments \cite{kaumanns_multipass_2018,russbueldt_scalable_2019} and was studied explicitly in a simulation paper \cite{hanna_hybrid_2020}. First signatures could be detected approximately 50\,dB below the maximal power spectral density for 30\,$\mu$J output pulse energy. At maximum power, the amplitude of this short wavelength feature becomes one order of magnitude stronger. We note that the mixed frequencies lie outside of the reflection band of our dielectric optics, and thus the measured spectrum does not represent the magnitude and central frequency of the side-bands generated inside the MPC. Our simulation also predicts spectral components around 1125\,nm (see Fig.~\ref{fig:S_FWM}(b)) which are outside of our detection range and could not be observed. \par 
 \begin{figure}[t]
 	\centering
 	\includegraphics[width=.7\linewidth]{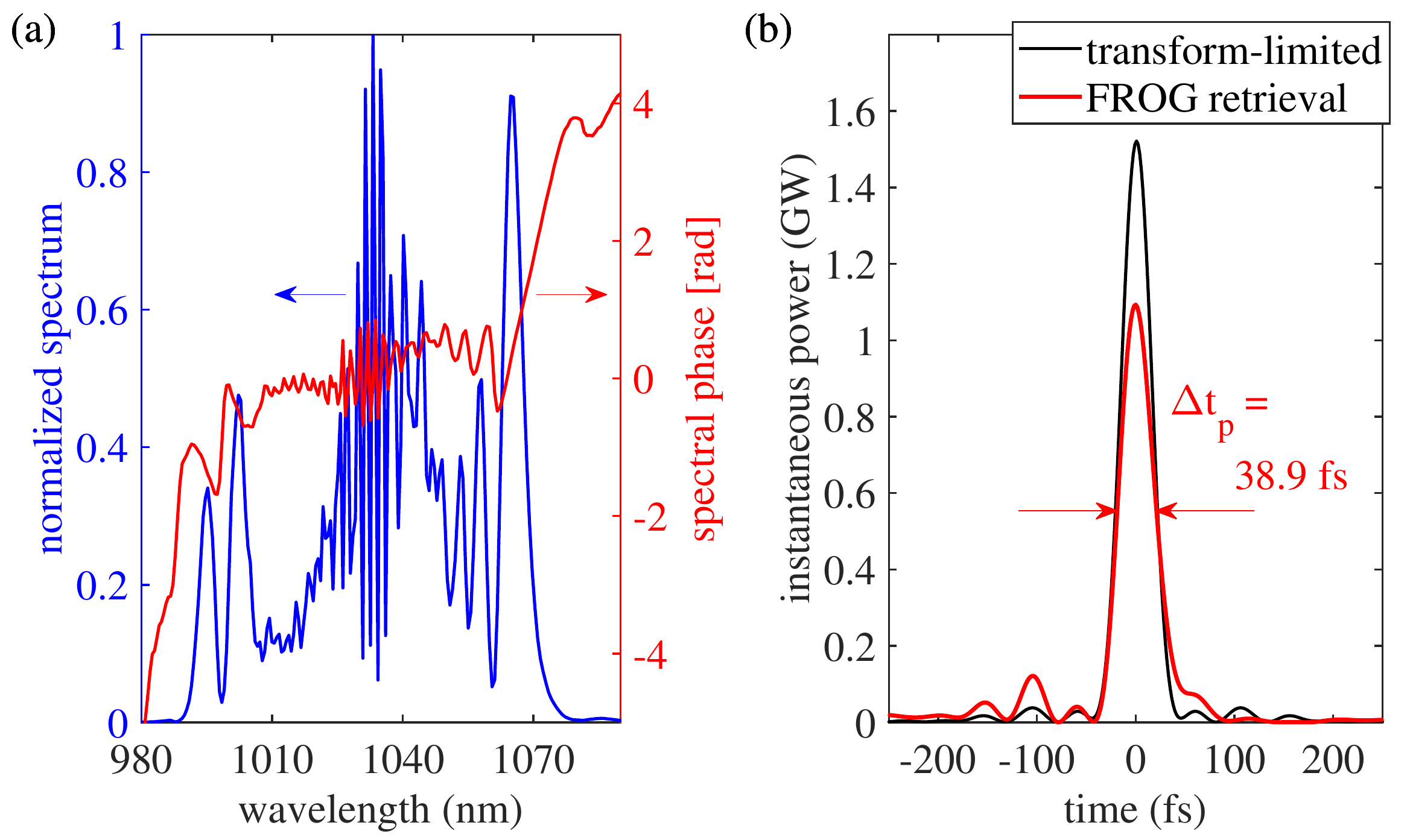}
 	\caption{\textbf{(a)} Retrieved FROG spectrum and spectral phase of the post-compressed pulses. \textbf{(b)} Retrieved pulse shape compared to the transform limited pulse for 65\,$\mu$J pulse energy.}
 	\label{fig:Fig5_FROG}
 \end{figure}
Despite the large spectral broadening factor, the measured spectrum was highly stable as Fig.~\ref{fig:Fig4_Broadening}(d) demonstrates. From one recording per second in an overnight measurement ($\approx$17 hours measurement time), we computed a standard deviation of the Fourier-transform limited pulse duration of only 0.1\,fs at 35.7\,fs mean value. The pulse-to-pulse energy fluctuations after the MPC exhibited only a 0.19\,\% standard deviation relative to the mean output energy. The measurements were performed with a single pulse per burst. Spectrum and pulse energy were logged in parallel.\par
To shorten the spectrally broadened pulses, we set-up a chirped-mirror compressor inducing a group delay dispersion (GDD) of -200\,fs$^2$ per reflection. For best compression, we changed the number of bounces in steps of four and slightly varied the input pulse energy to the MPC. By means of second harmonic FROG, we could at best retrieve a FWHM duration of 39\,fs (Fig.~\ref{fig:Fig5_FROG}), corresponding to a factor 32 pulse duration shortening in comparison to the 1.24\,ps input shown in supplement~\ref{S:FROG}. The required 72 bounces off the chirped mirrors applied a total GDD of -14\,400\,fs$^2$ to the pulses and resulted in only 78\,\% transmission of the compressor. Moreover, we measured 5\,\% depolarization after waveplate WP2. This is caused by the large nonlinear phase shift in the MPC which effectively amplifies cross-polarized light. Therefore, the usable pulse energy accumulated  to 65\,$\mu$J which yielded a peak power of 1.1 GW. By contrast to other bulk-MPC or multi-plate experiments \cite{fritsch_all-solid-state_2018,vicentini_nonlinear_2020,tsai_nonlinear_2019}, we did not observe a phase kink near the fundamental wavelength (Fig.~\ref{fig:Fig5_FROG}(a)).\par

\begin{figure}[t]
	\centering
	\includegraphics[width=\linewidth]{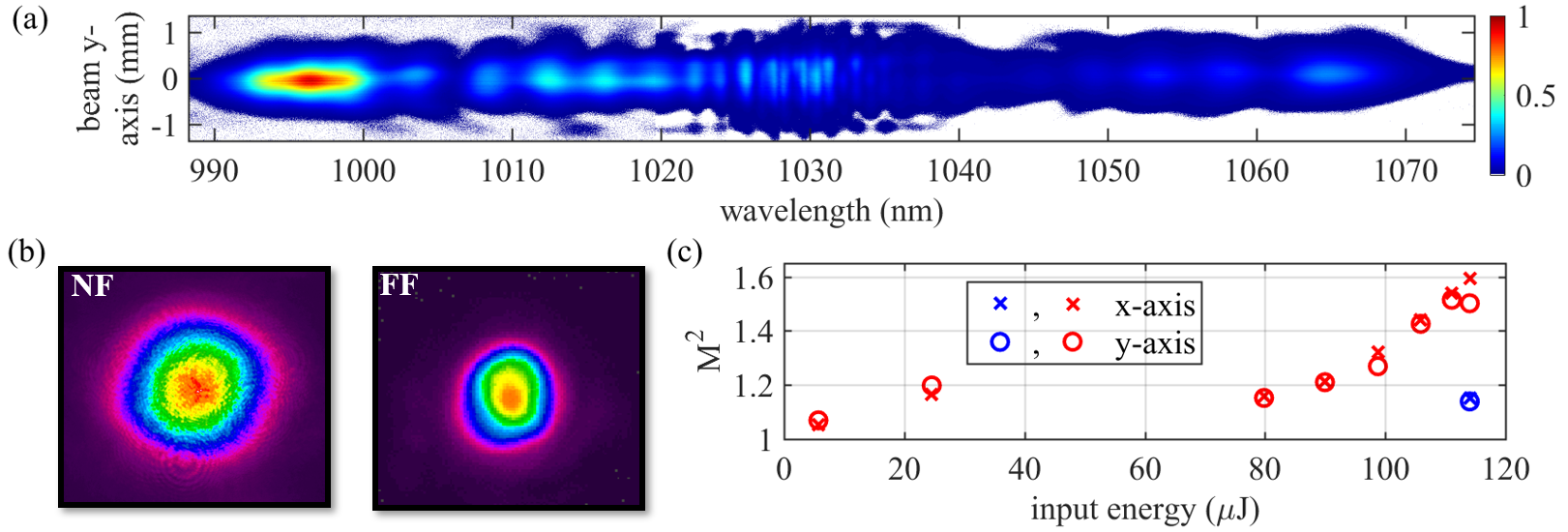}
	\caption{\textbf{(a)} Measured spectra across the vertical beam axis for a 36\,fs Fourier transform-limit. \textbf{(b)} Near-field (NF) and far-field (FF) profiles, also recorded at 36\,fs Fourier transform-limit. \textbf{(c)} Experimentally determined $M^2$ parameters for 32 roundtrips (red markers) and 8 roundtrips (blue markers).}
	\label{fig:Fig6_beam}
\end{figure}
\subsection{Beam characterization}
Fig.~\ref{fig:Fig6_beam}(a) shows a camera image after the $4f$-spectrograph at full spectral broadening. The frequency content is nearly uniformly distributed across the beam axis. We calculated a mean spectral homogeneity along the vertical beam axis of $\bar{V}_y = 97.3$\,\%, following the figure of merit proposed by Weitenberg et al. \cite{weitenberg_multi-pass-cell-based_2017} (see supplement~\ref{S:homogene}).  A homogeneity of $\bar{V}_y = 97.3$\,\% was also measured for a gas-filled MPC with comparable spectral broadening from 1.3\,ps to 39\,fs bandwidth limit \cite{kaumanns_multipass_2018}. This underscores that the detrimental spatial effects, which are typical for single-pass spectral broadening in the critical self-focusing regime, have been largely suppressed by the multi-plate multi-pass approach. \par 
Although the near- and far-field beam profiles for the highest spectral broadening factor displayed in Fig.~\ref{fig:Fig6_beam}(b) look close to Gaussian, we determined a distinct increase of the $M^2$ parameters to $1.6 \times 1.5$ at full input power. Fig~\ref{fig:Fig6_beam}(c) shows that we measured $M^2$ values of about 1.2 for all input energies up to 90\,$\mu$J, corresponding to a spectral broadening factor of about 30. However, for higher input energies, a gradual decrease of beam quality was recorded. A similar observation was reported in the first MPC paper by Schulte et al. \cite{schulte_nonlinear_2016} and attributed to thermal effects. I our work, we could run experiments at only 10\,Hz pulse repetition rate and can thus exclude thermal effects. In ref.~\cite{hanna_nonlinear_2017}, the degradation was interpreted as an onset of bulk-like broadening. This is, however, in contradiction to our compression and spectral homogeneity measurements. An experiment with the same input energy but considerably less roundtrips (blue markers Fig~\ref{fig:Fig6_beam}(c)) suggests that the higher $M^2$ values are caused by parasitic nonlinear effects in the MPC. We suspect that the four-wave mixing process (see supplement~\ref{S:FROG}, Fig.~\ref{fig:Fig4_Broadening}(c)) between broadened spectrum and unbroadened mode-mismatched pulse pedestal introduced non-periodic losses in the MPC which led to reshaping of the beam. Therefore, input pulse shaping could further improve the output beam quality.\par

\section{Discussion}
We have presented a hybrid bulk multi-pass, multi-plate method for spectral broadening of ultrashort laser pulses. This novel approach yielded significantly higher single-stage compression factors than previously reported from the individual techniques. In total, the laser beam passed 384 Kerr media in our setup. On the one hand, this is practically impossible to implement in a pure multi-plate scheme where less than ten passes are typical. On the other hand, to our knowledge all bulk-MPC papers roughly followed the initial proposal that the B-integral per pass through the Herriott-cell should not exceed $\pi/5$ \cite{andreas_vernaleken_method_2015}. The simulations shown in Fig.~\ref{fig:Fig3_Simulation} predict B-integrals per pass of up to $B\approx 4\pi/5$ resulting from an average peak irradiance $<I_p> = 285$\,GW/cm$^2$ inside the Kerr media. This was achieved through the introduction of multiple thin plates and explains why the presented 32-fold pulse duration reduction has not been accomplished with the bulk-MPC technique alone. It is only comparable to the best results demonstrated with gas-filled MPCs which operate in the sub-crititcal self-focusing regime \cite{kaumanns_multipass_2018,balla_postcompression_2020,kaumanns_spectral_2021}. By contrast to merely increasing the number of roundtrips in the MPC to attain large nonlinear phase shifts, the nonlinear mode-matching technique minimizes peak intensity variations in the Kerr media, and  thus enables large spectral broadening factors by high intensities and not primarily by long propagation lengths in the nonlinear material. This is essential if dispersion induces pulse stretching, i.e. for generating ultrashort pulses. Despite the large single-pass B-integrals and operating approximately 20 times above the critical power of the nonlinear material, we could vastly suppress undesired features of bulk spectral broadening, such as, spectral inhomogeneity across the beam profile and the typical phase kink near the input wavelength. This was enabled by distributing the Kerr nonlinearity along the MPC instead of using one thick fused silica window. The demonstrated approach was implemented with standard quarter-wave stack mirrors and off-the-shelf Kerr media. Neither a vacuum or high-pressure compatible enclosure, nor dispersive MPC mirrors were needed. Moreover, the setup was very compact and excellent stability was demonstrated. This makes the method highly attractive for turning a high-power picosecond laser into an ultrafast light source generating sub-50\,fs pulses. Whereas we measured constantly good beam quality up to a spectral broadening factor of 30, the $M^2$ value increased from 1.2 to 1.6 upon reducing the Fourier-transform limit further from 40\,fs to 36\,fs. We attribute this to parasitic four-wave mixing in the MPC which, according to our simulations, could be circumvented by pedestal-free input pulses.\par 
The main obstacle for generating even shorter pulses with our approach is the bandwidth of the used optics. The MPC mirrors support durations of about 30\,fs. Replacing tantalum pentoxide by titania as high index material of the quarter-wave stack mirrors would enable the generation of about 20\,fs pulses. By using enhanced metallic mirrors, even few-cycle pulses were demonstrated \cite{balla_postcompression_2020,muller_multipass_2021}. Conventional anti-reflection coatings would lead to clearly lower MPC transmissions if larger bandwidths with the same number of roundtrips are targeted. However, nanotextured surfaces exhibit very low Fresnel reflections over a large spectral range \cite{hobbs_continued_2012}. This opens the perspective of few-cycle pulse generation by the hybrid multi-pass multi-plate approach. Moreover, the Kerr media can be moved closer to the MPC mirrors if laser peak power increases \cite{schulte_nonlinear_2016}, and thus the proposed scheme can be employed in the 5\,MW to 5\,GW range without the need for Herriott-cells exceeding usual optical table lengths. For certain high peak power applications, bulk-MPCs may present interesting alternatives to gas-filled MPCs. Gas ionization and mirror damage thresholds differ by orders of magnitudes. Therefore, gas-filled MPCs are usually operated near the stability edge. In contrast, bulk Kerr media and dielectric mirrors exhibit comparable damage thresholds. Consequently, very compact Herriott-cells that do not compromise spectral broadening are possible. Bulk spectral broadening is also envisioned for PW class lasers \cite{khazanov_nonlinear_2019}. Similar to high-power multi-pass amplifier research, where thermal lenses need to be considered, power-scaling of the bulk-MPC approach would require to find a compact resonator design that can handle Kerr lensing. A possible solution is the use of folded bow-tie cavities \cite{heyl_high-energy_2021}.\par 
In conclusion, the hybrid multi-pass, multi-plate approach has enabled record-high single-stage pulse compression factors for these bulk spectral broadening methods. The demonstrated efficiency, beam and pulse quality is comparable to less compact and more complex multi-stage setups. In contrast to gas-filled MPCs, no sealed enclosure of the setup is required and our approach can be applied to a large peak power range. Therefore, it is a very attractive method for pulse-compression of state-of-the-art laser systems and opens perspectives for novel high repetition rate extreme light sources operating in the few-cycle or TW to PW peak power regime.\par

\section*{Acknowledgments}
We thank Caterina Vidoli for proof-reading the manuscript and we acknowledge DESY (Hamburg, Germany), a member of the Helmholtz Association HGF, for the provision of experimental facilities.

\subsection*{Author Contributions} 
M. Seidel conceived the hybrid spectral broadening scheme, conducted experiments as well as simulations and led the writing of the manuscript. P. Balla implemented the initial setup and conducted first experiments. C. Li and L. Winkelmann aligned the laser used for the experiments. G. Arisholm developed the SISYFOS code and adapted it to MPC simulations. L. Winkelmann, I. Hartl and C. M. Heyl initiated the project. I. Hartl and C.M. Heyl conceived the initial setup and supervised the project. All authors revised the manuscript and contributed to the writing of it.

\subsection*{Conflicts of Interest}
The authors declare that there is no conflict of interest regarding the publication of this article.

\subsection*{Data Availability}
Data can be obtained from the corresponding author upon reasonable request.

\newpage

\section*{Supplementary Materials}
\renewcommand{\thefigure}{S\arabic{figure}}
\renewcommand{\thesection}{S\arabic{section}}
\renewcommand{\thetable}{S\arabic{table}}
\setcounter{section}{0}
\setcounter{figure}{0}

\section{Simulated single pass enhancements}
\label{S:single_pass_enhancement}
We used the SISYFOS package \cite{arisholm_simulation_2012} to simulate nonlinear propagation of an intense Gaussian beam through fused silica and compared the pulse peak irradiance at the beginning and at the end of propagation. We varied propagation length, i.e. Kerr medium thickness, pulse energy and input beam size while keeping a fixed spectral broadening factor of about 1.3 (Table~\ref{tab:irr_enhancement}).  Whereas the irradiance enhancement within the nonlinear medium for 150\,$\mu$J pulse energy and a 1\,mm long plate was about 2\,\%, the enhancement was more than three times larger for a 3 mm long Kerr medium pumped at lower intensity. This is caused by self-focusing which leads to a highly nonlinear increase of on-axis irradiance in extended Kerr media if the critical power is exceeded \cite{seidel_all_2016}.

\begin{table}[b!]
	\caption{Simulation of irradiance enhancement inside a Kerr medium for a fixed broadening factor of 1.3 and Gaussian input pulses with $\Delta t_p = 750$\,fs duration.}
	\centering
	\begin{tabular}{ccccc}
		\hline
		B-integral & $l$ & $w$ & $E_p$ & $\mathcal{E}$ \\
		\hline
		$0.28\pi$ & 1\,mm & 117\,$\mu$m & 100\,$\mu$J & 3.4\,\% \\
		$0.27\pi$ & 1\,mm & 147\,$\mu$m & 150\,$\mu$J & 2.1\,\% \\
		$0.27\pi$ & 3\,mm & 147\,$\mu$m & 50\,$\mu$J & 6.5\,\% \\
		\hline
	\end{tabular}
	\caption*{$l$: Kerr medium length, $w$: $1/e^2$ beam radius, $E_p$: pulse energy, $\mathcal{E}$: irradiance enhancement.}
	\label{tab:irr_enhancement}
\end{table}

\section{Nonlinear mode-matching}
\label{S:mode-match}
A Herriott-type cell is usually operated as a $q$-parameter preserving resonator \cite{herriott_off-axis_1964,sennaroglu_design_2003}.  It, therefore, must satisfy the following ABCD matrix relation for Gaussian beams:

\begin{align}
	q &\stackrel{!}{=}  \frac{\tilde{M}_\text{MPC}(1,1)q+\tilde{M}_\text{MPC}(1,2)}{\tilde{M}_\text{MPC}(2,1)q+\tilde{M}_\text{MPC}(2,2)}, \label{eq:q-pres}\\ 
	&\text{where } \tilde{M}^\text{lin}_\text{MPC} =\tilde{L}(f)\tilde{P}(d).
	\label{eq:ABCDcold}
\end{align}
$\tilde{L}$ denotes the thin lens ABCD matrix, $f$ the MPC mirror focal length, $ \tilde{P}$ is the propagation ABCD matrix and $d$ the cell length which must be long enough to prevent optics damage. The beam radius of curvature (RoC) must be $2f$ at both MPC mirror planes for matching the q-parameter to the eigenmode of the MPC \cite{herriott_off-axis_1964,herriott_folded_1965}. The beam size for mode-matching is inferred from the cell length. Eq.~\eqref{eq:ABCDcold} holds for a passive cavity or linear propagation. Kerr lensing, however, modifies the ABCD matrix to: 
\begin{align}
	\tilde{M}_\text{MPC}^\text{nl} = \tilde{L}(f)\tilde{P}(d_{MK})\left[\tilde{K}(n,l,\gamma)\tilde{P}(d_K)\right]^{N-1}\tilde{K}(n,l,\gamma)\tilde{P}(d_{MK}),
	\label{eq:ABCDhot}
\end{align}
where $d_{MK}$ is the distance from the MPC mirror to the outer Kerr medium, $\tilde{K}$ is the Kerr lens ABCD matrix depending on the refractive index $n$, the plate length $l$, and the parameter $\gamma$, $d_K$ the plate spacing and $N$ the number of Kerr media. Only symmetric cells are considered. Matching again the RoCs of beam and mirrors at the mirrors leads to smaller spot sizes on the mirrors due to the additional lenses inside the MPC. We used the Kerr lens ABCD matrix derived in ref.~\cite{siegman_lasers_1986} for a quadratic duct and added a scaling parameter $\kappa$ to account for the Gaussian beam shape \cite{li_yan_pulse_1994}:
\begin{align}
	\tilde{K}(l,\gamma) &= \begin{pmatrix} \cos (\gamma l) & \sin (\gamma l)/(n\gamma) \\ n\gamma\sin (\gamma l) & \cos (\gamma l)  \end{pmatrix}, \label{eq:ABCD_Kerr} \\
	\gamma &= \frac{\kappa}{w^2}\left(\frac{8n_2P_p}{\pi n}\right)^{1/2}, \label{eq:gamma}
\end{align}
where $w$ is the beam radius, $n_2$ the nonlinear refractive index and $P_p$ the pulse peak power. After having fixed the nonlinear refractive index $n_2^\text{sim}$ (see section~\ref{sec:meth_sim}), the $\kappa$ parameter was derived from a Z-scan-type simulation resembling the setup shown in Fig.~\ref{fig:S_Zscan}(a). The Gaussian beam width at the camera plane was computed for eight Kerr-lens positions and fitted by \eqref{eq:ABCD_Kerr}. The comparison between the best-fit $\gamma$ from \eqref{eq:ABCD_Kerr} and the analytical $\gamma$ from \eqref{eq:gamma} yielded $\kappa = 0.43$.\par
\begin{figure}[b!]
	\centering
	\includegraphics[width=.7\linewidth]{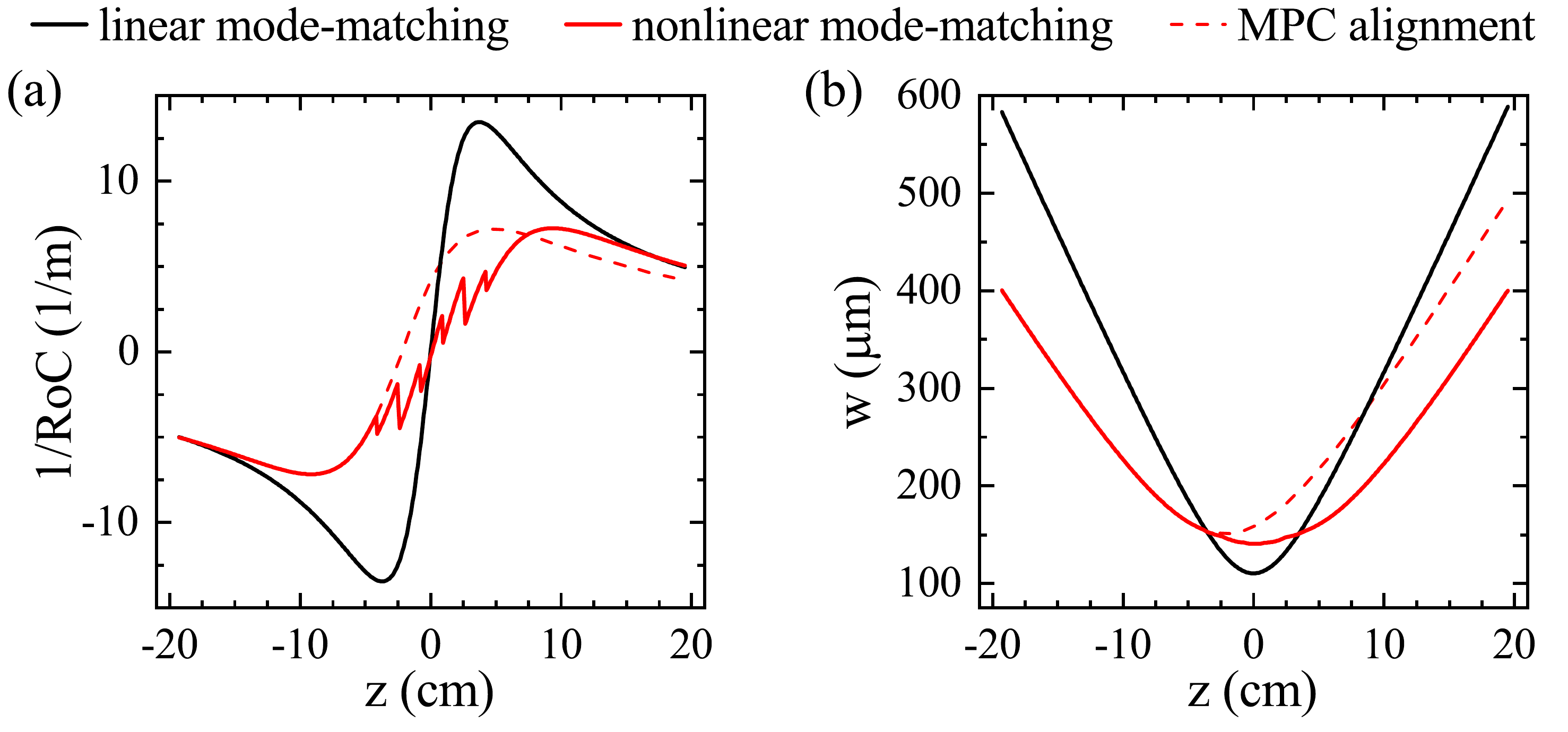}
	\caption{\textbf{(a)} Inverse radius of curvature ($1/\text{RoC}$) and \textbf{(b)} beam radius (w) for a mode-matched MPC without nonlinear media (black solid lines) and with six Kerr lenses (red solid lines). Only the radii of curvature at the mirrors are identical. Otherwise, beam sizes and curvatures differ considerably during a single pass. The red dashed lines show inverse beam curvature and beam radius for a nonlinear mode-matching setting and a weak beam, i.e. negligible Kerr lensing. This configuration was used to align the MPC.}
	\label{fig:S_mode-match}
\end{figure}
Fig.~\ref{fig:S_mode-match} shows an example of single-pass ABCD matrix calculations for a cold Herriott-cell where linear mode-matching applies and the MPC with six Kerr media that we experimentally studied. In both cases, the radii of curvature at the mirrors are $-5$\,m after and $+5$\,m before reflection. Otherwise, the q-parameters evolve very differently inside the MPC. The linearly mode-matched beam is about 50\,\% larger at the mirrors than the nonlinearly mode-matched beam. This does not cause damage issues since the damage threshold of the Kerr media is critical. The focus spot size of the nonlinearly mode-matched MPC is larger than that of the linearly mode-matched cell. This is different from nonlinear mode-matching in a gas-filled cell where the beam divergence is effectively reduced by the gas nonlinearity \cite{hanna_nonlinear_2021}. By contrast, a weak quasi-guiding structure is observable in Fig.~\ref{fig:S_mode-match}. The slope of the radius of curvature changes significantly at every Kerr medium resulting in much smaller beam size variations near the center of the MPC than in the linearly mode-matched case that exhibits usual Gaussian beam diffraction.\par
To align the MPC in our experiments, we calculated the beam parameter $q_\text{M}$ on the cavity mirror that fulfilled Eqs.~\eqref{eq:q-pres} and \eqref{eq:ABCDhot} for the maximal input energy of 112\,$\mu$J. Subsequently, we computed the beam path for linear propagation (red dashed lines in Fig.~\ref{fig:S_mode-match}) since alignment was done at low input power. By placing a beam profiler into the predicted focal plane and varying the distance between the lenses L1 and L2 (see Fig.~\ref{fig:Fig2_Setup}), as well as the position of the translation stage with the mirrors M2 and M3, the measured beam waist was match to our calculations. Afterward, the silica plates were inserted and the energy was ramped up to 112\,$\mu$J without further modifications of the mode-matching optics.

\begin{figure}[b!]
	\centering
	\includegraphics[width=.19\linewidth]{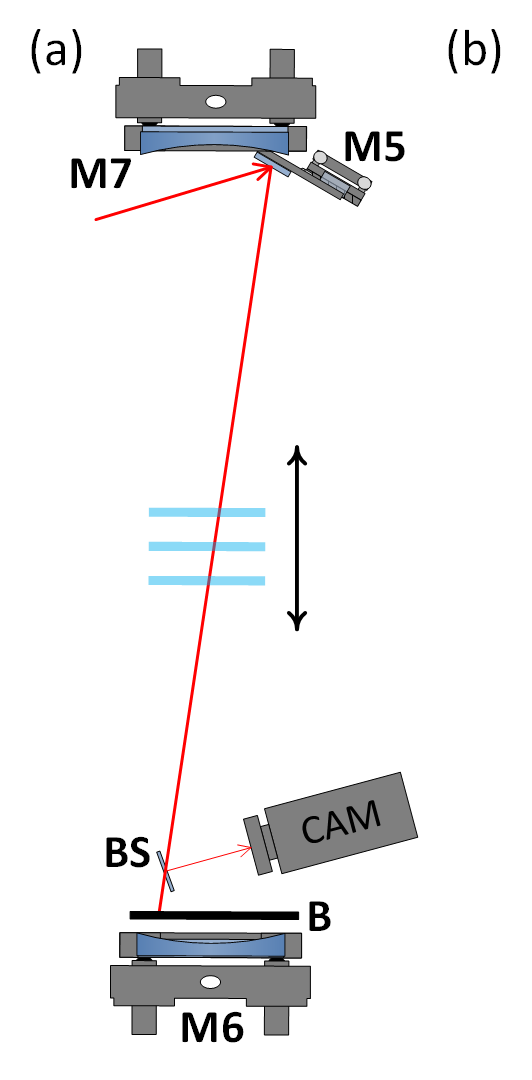} \hspace*{-5mm}
	\raisebox{2mm}{\includegraphics[width=.41\linewidth]{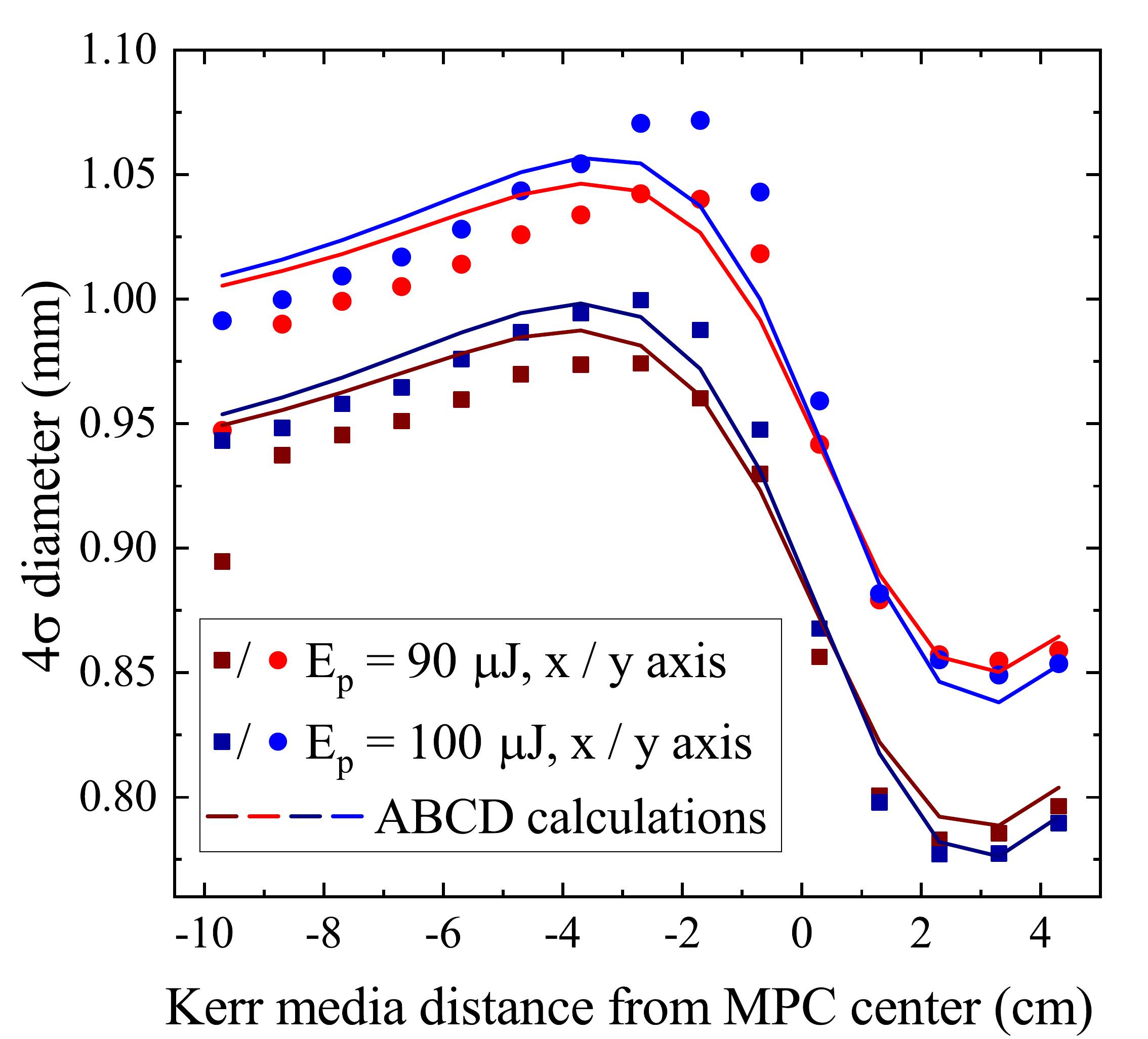}}
	\caption{\textbf{(a)} The Z-scan experiments were done in the MPC arrangement shown in Fig.~\ref{fig:Fig2_Setup} using all optics up to mirror M5. A beam sample BS was inserted and the beam was blocked (B) in front of the curved mirror M6. The beam sizes were measured with the camera (CAM)  \textbf{(b)} Measured beam diameters in x- and y-direction for 90\,$\mu$J and 100\,$\mu$J input pulse energies. The spot sizes were fitted by applying Eqs.~\eqref{eq:ABCDhot}-\eqref{eq:gamma} and optimizing $n_2^\text{exp}$ (solid lines). The observation of varying beam sizes manifest the impact of Kerr-lensing on the beam inside the MPC.}
	\label{fig:S_Zscan}
\end{figure}
\begin{figure}[b!]
	\centering
	\includegraphics[width=.7\linewidth]{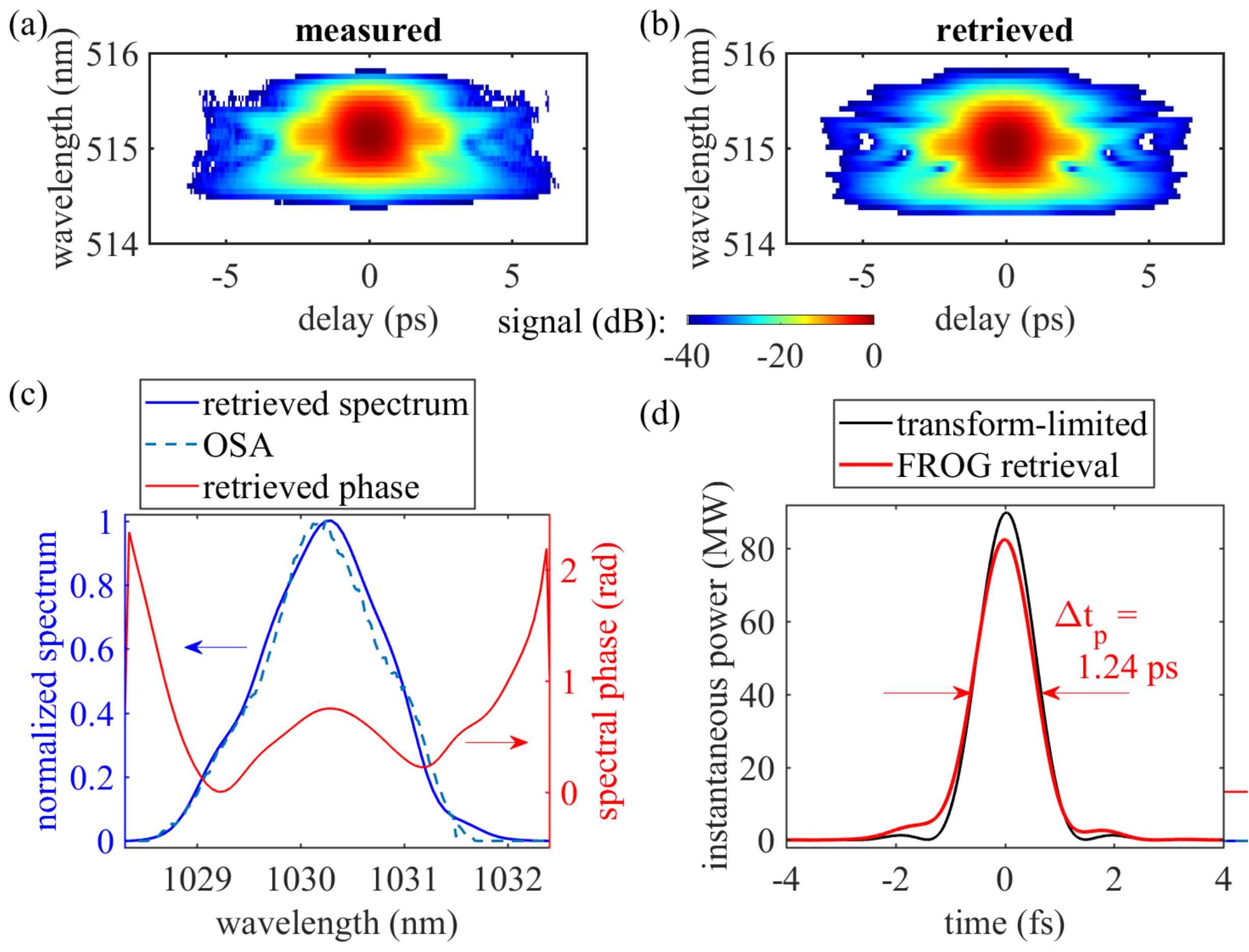}
	\caption{\textbf{(a)} Measured and \textbf{(b)} retrieved FROG traces of the input pulses in logarithmic scale. The grid size was 256\,x\,256 points and the FROG error 0.2\,\%. \textbf{(c)} Left axis: Retrieved spectrum (solid blue line) and spectrum recorded with an optical spectrum analyzer (dashed light blue line, OSA). Right axis: Retrieved spectral phase (red line). \textbf{(d)} Retrieved pulse (red line) and Fourier transform-limited pulse (black line) assuming 112$\,\mu$J pulse energy.}
	\label{fig:S_FROG1240fs}
\end{figure}

\section{Z-scans}
\label{S:Z-scan}
Before conducting spectral broadening experiments, we measured an effective $n_2^\text{exp}$ by a Z-scan type measurement.  This was done because of the beam shape dependence of the $\gamma$ parameter and the uncertainty in the determination of the peak power $P_p$. We moved a set of three Kerr media along the MPC shown in Fig.~\ref{fig:Fig2_Setup} and measured the beam size variations at the MPC mirror M6 plane (see Fig.~\ref{fig:S_Zscan}(a)). By the ABCD matrix calculus from Eqs.~\eqref{eq:ABCDhot}-\eqref{eq:gamma}, we determined $n^\text{exp}_2 = 2.25\cdot10^{-16}$cm$^2$/W. The calculated spot sizes reproduced the measured ones well as Fig.~\ref{fig:S_Zscan}(b) exhibits. The derived nonlinear refractive index is in good agreement with the used simulation and the literature values \cite{milam_review_1998}. 

\begin{figure}[t]
	\centering
	\includegraphics[width=.6\linewidth]{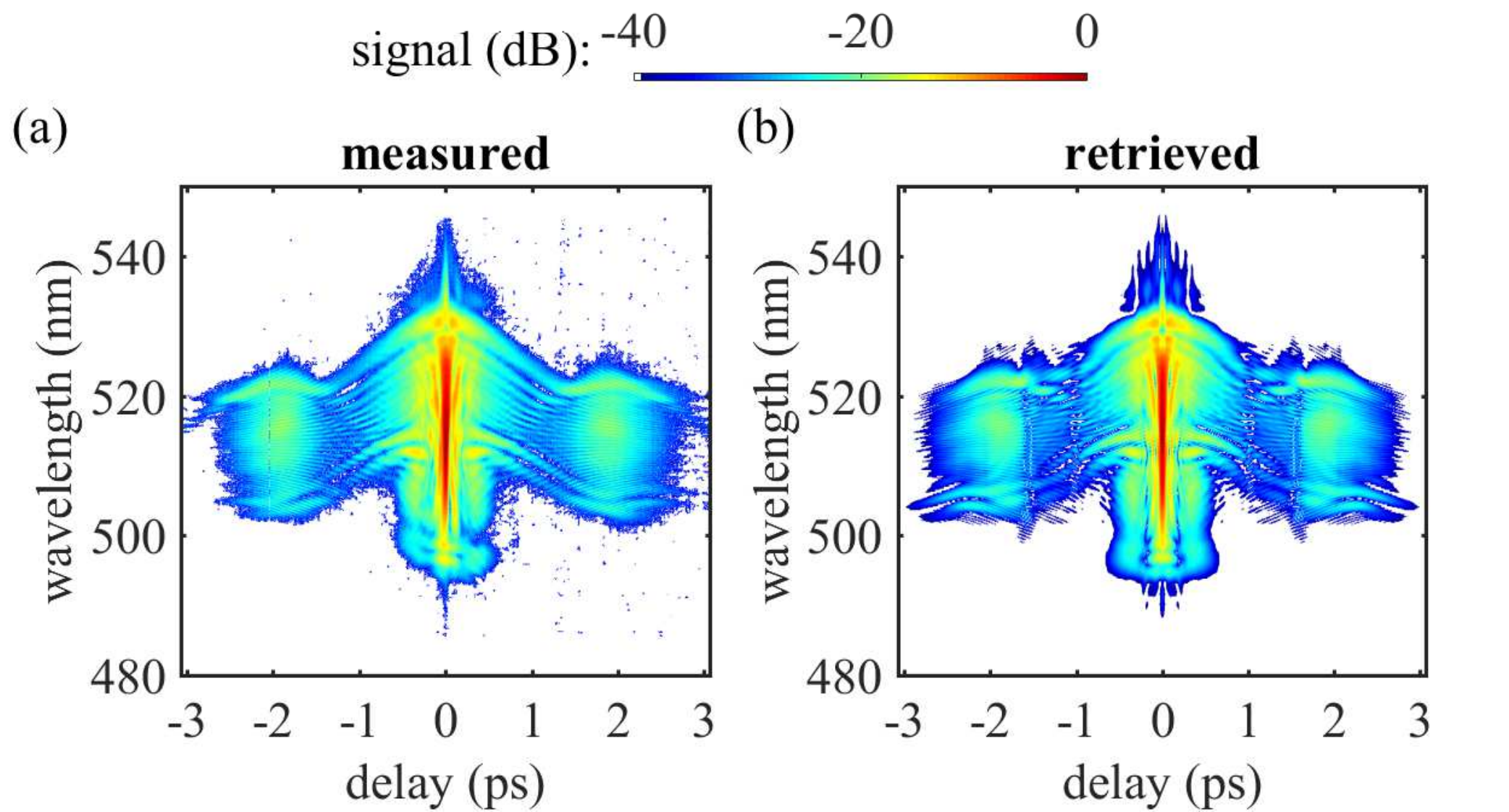}\\
	\includegraphics[width=.5\linewidth]{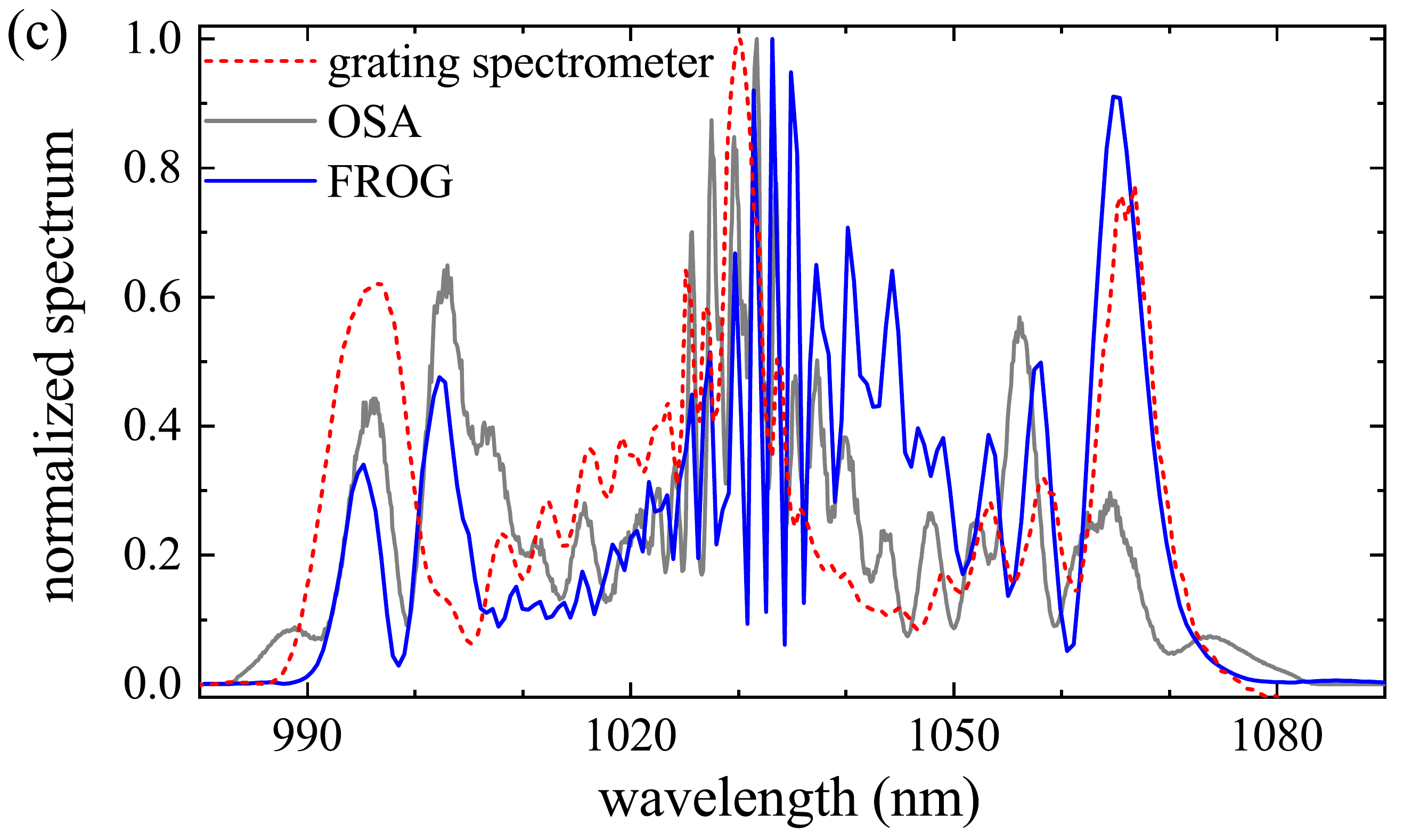}
	\caption{\textbf{(a)} Measured and \textbf{(b)} retrieved FROG traces in logarithmic scale. The grid size was 1024\,x\,1024 points and the FROG error 0.4\,\%. \textbf{(c)} Comparison of spectra corresponding to the compressed pulse retrieval. We measured with a compact grating spectrometer (red dashed line) as well as with an optical spectrum analyzer (gray solid line, OSA). As in Fig.~\ref{fig:Fig5_FROG}, the blue solid line shows the spectrum derived from the FROG trace.}
	\label{fig:S_FROG39fs}
\end{figure}
\begin{figure}[t]
	\centering
	\includegraphics[width=.7\linewidth]{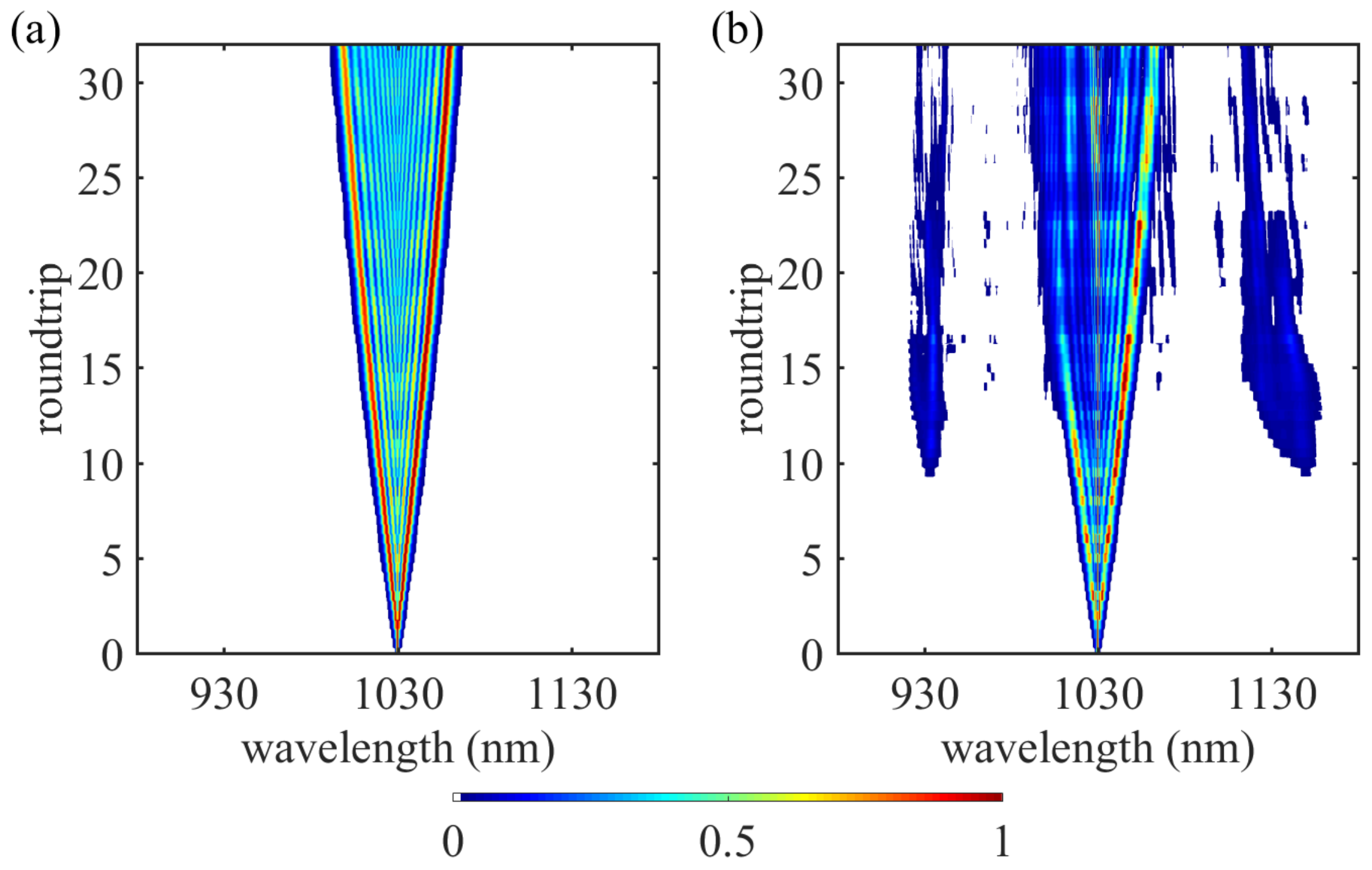}
	\caption{\textbf{(a)} Simulated Evolution of spectral broadening for Gaussian input pulses (see Fig.~\ref{fig:Fig3_Simulation}) in comparison to \textbf{(b)} input pulses with the shape of Fig.~\ref{fig:Fig2_Setup}(b) and the same peak power. The latter case shows clear signatures of parasitic four-wave mixing. The spectra were normalized at each propagation length.}
	\label{fig:S_FWM}
\end{figure}
\section{FROG measurements}
\label{S:FROG}
To determine the experimental compression factor and to understand the generation of the spectral side-bands shown in Fig.~\ref{fig:Fig4_Broadening}(d), we measured the input pulse shape by FROG. The results are shown in Fig.~\ref{fig:S_FROG1240fs}. We retrieved a nearly transform-limited pulse of 1.24\,ps duration which, however, exhibits a pedestal, i.e., a non-Gaussian shape. For the maximum pulse energy of 112\,$\mu$J, the pulse peak power was about 80\,MW.\par 
Moreover, we provide additional information about the FROG measurement resulting in the 39\,fs pulses displayed in Fig.~\ref{fig:Fig5_FROG}. The measured and retrieved FROG traces as well as a comparison between measured and retrieved spectra are shown in Fig.~\ref{fig:S_FROG39fs}. Although the FROG error was 0.4\,\%, the fundamental spectrum measured with an optical spectrum analyzer (OSA) only reproduced the wiggles of the FROG spectrum well, whereas the spectral amplitudes differed. We note that there are additional side lobes around 990 nm and 1070 nm visible in the OSA spectrum which are neither present in the retrieved FROG spectrum, nor in the one measured with a grating spectrometer. These might emerge from additional nonlinearity accumulated by steering the beam into the OSA. Due to the burst-mode operation of our laser, relatively high pulse energies are needed to measure a spectrum with an OSA. On the other hand, our silicon-based compact grating spectrometer cannot resolve the fast oscillations near the fundamental wavelength. We note that the combination of the spectra shown in Fig.~\ref{fig:S_FROG39fs}(c) with the retrieved phase from Fig.~\ref{fig:Fig5_FROG}(a) results only in a pulse duration variation from 36\,fs to 39\,fs.

\section{Parasitic four-wave mixing}
\label{S:FWM}
We simulated the spectrum evolution inside the MPC for Gaussian pulses and the retrieved pulses shown in Fig.~\ref{fig:S_FROG1240fs}(d). Whereas the spectral broadening of Gaussian pulses, visualized in Fig.~\ref{fig:S_FWM}(a), resembles one-dimensional textbook examples for self-phase-modulation \cite{agrawal_self-phase_2013}, simulations with the experimentally measured input pulses result in more power near the fundamental wavelength, a pronounced peak at the long-wave edge of the self-phase modulated spectrum and, most striking,  the generation spectral side-bands. These resemble the experimental observation shown in Fig.~\ref{fig:Fig4_Broadening}(d).

\section{Homogeneity calculation}
\label{S:homogene}
We calculated the figure of merit proposed by Weitenberg et al. \cite{weitenberg_multi-pass-cell-based_2017} in the following way:
\begin{align}
	V_y(y_i) &= \left[\sum_j\left\{I(\lambda_j,y_i)I_0(\lambda_j)\right\}^{1/2}\right]^2 \left[\sum_j I(\lambda_j,y_i)\sum_k I_0(\lambda_k)\right]^{-1},\\
	\bar{V}_y &= \sum_i V_y(y_i) P(y_i) \left[ \sum_j P(y_j) \right]^{-1},
	\label{eq:beam_hom}
\end{align}
where $\lambda_j$ are the camera pixels in x-direction representing wavelength, $y_i$ are the camera pixels in y-direction representing the vertical beam axis, $I(\lambda_j,y_i)$ are the camera counts, $I_0(\lambda_j)$ the on-axis counts and $P(y_i) = \sum_j I(\lambda_j,y_i)$ is the vertical beam profile. We summed over the $6\sigma$ width of $P(y_i)$ and over the full width at 0.2\,\% of the maximum of the average spectrum. The level corresponded to our signal to noise ratio. Due to the isotropy of all optics, we did not explicitly measure the homogeneity in x-direction. 

\newpage

\printbibliography

\end{document}